\documentclass[11pt]{article}
\usepackage{latexsym}
\usepackage{amsfonts}

\newtheorem{theorem}{Theorem}
\newtheorem{lemma}{Lemma}

\begin{document}
\title{On global existence for the spherically symmetric
  Einstein-Vlasov system in Schwarzschild coordinates} 
\author{H\aa kan Andr\'{e}asson} 
\maketitle

\begin{abstract} 
The spherically symmetric Einstein-Vlasov system in Schwarzschild
coordinates (i.e. polar slicing and areal radial coordinate) 
is considered. An improved continuation criterion
for global existence of classical solutions is given. Two other types
of criteria which prevent finite time blow-up are also given. 
\end{abstract} 

\section{Introduction} 
A central issue in general relativity is the problem of spherically
symmetric gravitational 
collapse and to determine whether or
not the weak- and strong cosmic censorship conjectures (CCC) hold true
in this case. These conjectures were raised by Penrose in the late
'60s. By now they have precise mathematical
formulations~\cite{Cu4} and are considered to be among the most
important problems in classical general relativity. In brief the
meaning of the weak CCC is as follows: The Penrose 
singularity theorem says that there are initial data that will lead to
singular spacetimes, i.e. geodesically incomplete spacetimes. The weak 
cosmic censorship states that generically, singularities will always be
covered by an event horizon and distant observers will never be able
to ``see'' the singularities. Singularities that can be ``seen'' are
called naked. To answer the CCC 
a detailed picture of the global properties of the solutions to 
the Cauchy problem for the Einstein matter equations are 
necessary.

In 1999 Demetrios Christodoulou finished his long term project on
the spherically symmetric Einstein-Scalar Field system (ESF) by
proving 
both the weak- and strong cosmic censorship conjecture~\cite{Cu2} by a 
careful analysis of the Cauchy problem. 
This important work is the only case where such a study has been
completed in an affirmative sense. On the contrary, Christodoulou has
shown that when the matter model is
dust (i.e. a pressureless perfect fluid) then there are
naked singularities generically~\cite{Cu6}. It is however a rather
common belief that a dust model is not an appropriate matter model
for studying cosmic censorship since the pressure is assumed to be
zero. A study of a more general fluid model for gravitational
collapse suffers at present from a lack of satisfying mathematical
results of such equations in the absence of gravity and in the case of
Newtonian gravity. A successful treatment in these
cases seems necessary before it will be mathematically meaningful to
couple such a model to the Einstein equations. A matter model that is
known to have satisfying mathematical features is the Vlasov model
which gives a kinetic description of the matter. In particular it was
shown 
independently by Pfaffelmoser~\cite{Pf} and Lions and Perthame~\cite{LP}
that in the case of Newtonian gravity, where a kinetic description
gives rise to the Vlasov-Poisson system, classical solutions
exist globally in time. This system
can now be said to be well-understood mathematically and it is
therefore natural to choose the Vlasov model for investigating
gravitational collapse in general relativity. In particular this
matter 
model is phenomenological in contrast to a scalar field which is
said to be field theoretical. It would certainly be 
desirable to establish similar results for a phenomenological
model as Christodoulou has done for a scalar field.
One way to attack the cosmic censorship conjecture is to prove
global existence in a singularity avoiding coordinate system. It is
then often conjectured that cosmic censorship would follow 
straightforwardly. 
The first attempt by Christodoulou on the ESF 
system was indeed of this nature. However, in that case it turned out
that such an approach was doomed to failure~\cite{Cu4} by finding initial
data that lead to naked singularities~\cite{Cu1}. In the case of Vlasov
matter it is still an open question whether or not naked
singularities form for any initial data. (Note that cosmic censorship
holds true if the set of initial data which generates naked
singularities is a null set). Thus the approach of 
proving global existence in a certain coordinate system is yet a
possile path to weak CCC. The aim of this paper is to present some
new results on the problem of global existence of the spherically
symmetric Einstein-Vlasov system in Schwarzschild coordinates. In
particular we give an improved continuation criterion. We also
consider two other types of criteria, one in which we study the case
when all matter is assumed to fall inwards. A continuation criterion
for this system was first given by Rein and Rendall
in~\cite{RR1}. To discuss the content of this criterion we need to
introduce the main concept in kinetic theory, the distribution
function. Consider a collection of particles, where the particles are
stars, galaxies or even clusters of galaxies. A characteristic feature
of kinetic theory is that the model is 
statistical and that the particle system is described by a distribution
function $f=f(t,x,p)$, which represents the density of particles
with given space-time position
$(t,x)\in\mathbb{R}\times\mathbb{R}^3$ and momentum $p\in
\mathbb{R}^{3}$. 
Let us define the function $Q(t)$ by 
\begin{equation}
Q(t):=\sup \{|p|:\exists (s,x)\in [0,t]\times \mathbb{R}^3\mbox{ such that
    }f(s,x,p)\not= 0\}.\label{Q}
\end{equation} 
Hence, there are no particles at time $t$ having larger momentum than
$Q(t)$ in the system. If we now consider a solution of
the spherically symmetric Einstein-Vlasov system on a time interval
$[0,T[,$ then it is proved in~\cite{RR1} that if $Q(t)$ is bounded on
$[0,T[$ then the solution can be extended beyond $T.$ This is called a
continuation criterion. 
This is analogous to the situation for
the Vlasov-Maxwell system where the same condition ensures
that the solution can be extended. This was proved in a classical work
by Glassey and Strauss~\cite{GSt1} but recent results by
Pallard~\cite{P1},~\cite{P2} improve this criterion. 

The outline of the paper is as follows. The Einstein-Vlasov system in
Schwarzschild coordinates is introduced in section 2. A brief review
of previous results is also given here. In section 3 we present some general
a priori bounds not previously in the literature. In section 4 we make a 
comparison between the Einstein-Vlasov- and the Vlasov-Poisson system
to point out the main differencies. An improved 
continuation criterion is proved in section 5 using a mixed $L^{3+}$
and weighted $L^{\infty}$ approach. In section 6 an approach based on
hypotheses along characteristics is given. A discussion about
what is expected to happen if these bounds are not satisfied is also
included. In 
section 7 the case when all matter is falling inwards is studied. It
is seen that the well-known result that a Schwarzschild spacetime with
mass $M$ has closed null rays when $r=3M,$ also plays an essential role
for the Einstein-Vlasov system. 
\section{The Einstein-Vlasov system} 
For a derivation of the system given below we refer to~\cite{R}
and~\cite{RRS1}. 
In Schwarzschild coordinates the spherically symmetric metric takes
the form 
\begin{equation}
ds^{2}=-e^{2\mu(t,r)}dt^{2}+e^{2\lambda(t,r)}dr^{2}
+r^{2}(d\theta^{2}+\sin^{2}{\theta}d\varphi^{2}). 
\end{equation}
The Einstein equations are 
\begin{eqnarray} 
&\displaystyle e^{-2\lambda}(2r\lambda_{r}-1)+1=8\pi r^2\rho,&\label{ee1}\\ 
&\displaystyle e^{-2\lambda}(2r\mu_{r}+1)-1=8\pi r^2 p,&\label{ee2}\\ 
&\displaystyle\lambda_{t}=-4\pi re^{\lambda+\mu}j,&\label{ee3}\\ 
&\displaystyle e^{-2\lambda}(\mu_{rr}+(\mu_{r}-\lambda_{r})(\mu_{r}+
\frac{1}{r}))-e^{-2\mu}(\lambda_{tt}+\lambda_{t}(\lambda_{t}-\mu_{t}))=
4\pi q.&\label{ee4} 
\end{eqnarray} 
The Vlasov equation for the density distribution function
$f=f(t,r,w,F)$ reads 
\begin{equation} 
\partial_{t}f+e^{\mu-\lambda}\frac{w}{E}\partial_{r}f
-(\lambda_{t}w+e^{\mu-\lambda}\mu_{r}E-
e^{\mu-\lambda}\frac{F}{r^3E})\partial_{w}f=0,\label{vlasov} 
\end{equation} 
where 
\begin{equation}
E=E(r,w,F)=\sqrt{1+w^{2}+F/r^{2}}.\label{E} 
\end{equation}
The matter
quantities are defined by 
\begin{eqnarray} 
\rho(t,r)&=&\frac{\pi}{r^{2}}
\int_{-\infty}^{\infty}\int_{0}^{\infty}Ef(t,r,w,F)\;dwdF,\label{rho}\\ 
p(t,r)&=&\frac{\pi}{r^{2}}\int_{-\infty}^{\infty}\int_{0}^{\infty}
\frac{w^{2}}{E}f(t,r,w,F)\;d
wdF,\label{p}\\ 
j(t,r)&=&\frac{\pi}{r^{2}}
\int_{-\infty}^{\infty}\int_{0}^{\infty}wf(t,r,w,F),\;dwdF\label{j}\\ 
q(t,r)&=&\frac{\pi}{r^{4}}\int_{-\infty}^{\infty}\int_{0}^{\infty}\frac{F}{E}f(t,r,w,F)\;
dwdF.\label{q} 
\end{eqnarray} 
For technical reasons we also introduce the quantity 
\begin{equation}
k(t,r)=\frac{\pi}{r^{2}}\int_{-\infty}^{\infty}\int_{0}^{\infty}
|w|f(t,r,w,F)\;dwdF,\label{k} 
\end{equation} 
so that $k$ is nonnegative and dominates $p$ and $|j|$ but is
dominated by $\rho.$ 
The variables $w\in (-\infty,\infty)$ and $F\in [0,\infty)$ are the
radial component of the momentum and the square of
the angular momentum respectively. 
The following boundary conditions are imposed 
\begin{equation} 
\lim_{r\rightarrow\infty}\lambda(t,r)=\lim_{r\rightarrow\infty}\mu(t,r)
=\lambda(t,0)=0,\;\forall t\geq 0.\label{bdry} 
\end{equation} 
The last condition follows by requiring a regular center. 
We point out that the Einstein equations are not independent and that
e.g. the equations (\ref{ee3}) and (\ref{ee4}) follow by 
(\ref{ee1}), (\ref{ee2}) and (\ref{vlasov}). 

For an extensive review of previous results of this system see~\cite{An1}. 
Here we give a short summary of the results concerning the Cauchy problem. 
The issue of global existence for this system was first investigated
by Rein and Rendall in~\cite{RR1} where 
they obtain global existence for small (compactly supported) initial
data and in addition they show that the resulting spacetime 
is geodesically complete. They also prove the continuation
criterion mentioned in the introduction. In~\cite{Rl6} Rendall studies
the same system using a different set of coordinates (maximal
slicing and isotropic radial coordinate). Global existence
for small initial data is
obtained as well as an equivalent continuation criterion. In~\cite{Rl7}
Rendall shows that there exists initial data leading to trapped
surfaces and therefore to singular spacetimes by the Penrose
singularity theorem. The natural question is then if the weak CCC
holds. To investigate this it is necessary to consider 
large initial data. This 
is done in~\cite{RRS1} and~\cite{Rl6}. In the former case
Schwarzschild coordinates are used and it is proved that if matter
stays uniformly away from the center of symmetry then global existence
follows. In the latter case where the maximal and isotropic gauge is
used a similar result is shown assuming an additional condition on one
of the metric coefficients. The advantage being that the result is
obtained in a more direct way and the analysis of the Vlasov equation
is not necessary. 
Recently, Dafermos and Rendall~\cite{DR} have shown a similar result
for the Einstein-Vlasov system in double-null coordinates which has a
very interesting and important consequence. In the proof of the CCC
for the ESF system Christodoulou 
analyzes the formation of trapped surfaces and their
presence play an important role in his proof. Dafermos~\cite{D} then
strengthens the relation between trapped surfaces and weak CCC; it is
shown that a single trapped surface or marginally trapped surface in
the maximal development implies weak CCC. The theorem in~\cite{D}
requires certain hypotheses on the solutions of the Einstein matter
system. The main purpose of~\cite{DR} is to show that these hypotheses
are satisfied for the spherically symmetric Einstein-Vlasov system, 
which is indeed shown to be the case. This result might thus be useful
in the search for a proof of the weak CCC. 
Two numerical investigations have been
carried out where the primarily goal was to study critical collapse and
determine if Vlasov matter is of type I or type II, i.e. if arbitrary
small black holes can form (type II) or if there is a mass gap (type
I). In the first study~\cite{RRS2} Schwarzschild
coordinates were used and in the second~\cite{OC} maximal
slicing and an areal radial coordinate was used. Both of 
these studies indicate that Vlasov matter is of type I, and moreover,
no signs of singularity formation could be seen in either of these two
cases which support the idea that global existence holds in these
two coordinate systems. The investigation in this paper concerns the
Einstein-Vlasov system in Schwarzschild coordinates but we plan to
investigate other choices of gauge conditions both
analytically and numerically in the future. 

Let us now specify the sets of initial data that will be considered. 
Regular initial data is defined as the set of nonnegative compactly
supported smooth functions  
$f_{0}=f_{0}(r,w,F),$ such that 
\begin{displaymath} 
4\pi^{2}\int_{0}^{r}\int_{-\infty}^{\infty}\int_{0}^{\infty}
Ef_{0}(r,w,F)dwdFdr<\frac{r}{2}. 
\end{displaymath} 
The last condition says that no trapped surfaces are present 
initially. 
Given $R_{-}>0,$ 
we define the class of regular initial data with radial cut-off,
$I_{R_{-}},$ as the subset of regular initial data 
such that $f_{0}=0$ when $r\leq R_{-}.$ Likewise, given $F_{-}>0,$ the
subset $I_{F_{-}}$ such that $f_{0}=0$ when
$F\leq F_{-},$ 
is called regular initial data with angular momentum
cut-off, and consequently, $I_{R_{-},F_{-}}$ is the subset of regular
initial data with radial and angular momentum cut-off. Thus,
for initial data in $I_{R_{-},F_{-}}$ there is no matter in a 
ball around the centre and there is a lower bound on the angular
momentum $F.$ By conservation of angular momentum the latter property
is then true for all times. (Angular momentum cut-off has shown to 
be essential also in previous works, see~\cite{GSc4} and~\cite{ACR}.) 
We also make the definitions 
\begin{equation} 
F_{+}:=\sup\{ F:\exists (r,w)\in\mathbb{R}^{2} \mbox{ such that
  }f_{0}(r,w,F)\neq0\},\label{Fmax} 
\end{equation} 
and 
\begin{equation} 
R_{+}:=\inf\{R:f_{0}(r,\cdot,\cdot)=0\mbox{ for all }r\geq R\}.\label{Rmax} 
\end{equation}
Let us write down a couple of known 
facts about the system (\ref{ee1})-(\ref{bdry}). 
A solution to the Vlasov 
equation can be written 
\begin{equation}
f(t,r,w,F)=f_{0}(R(0,t,r,w,F),W(0,t,r,w,F),F), 
\label{repre} 
\end{equation} 
where $R$ and $W$ are solutions of the characteristic system
\begin{eqnarray}
\frac{dR}{ds}&=&e^{(\mu-\lambda)(s,R)}\frac{W}{E(R,W,F)},\label{char1}\\ 
\frac{dW}{ds}&=&-\lambda_{t}(s,R)W-e^{(\mu-\lambda)(s,R)}\mu_{r}(s,R)E(R,W,F)
\nonumber\\
&\phantom{hej}&+e^{(\mu-\lambda)(s,R)}\frac{F}{R^3E(R,W,F)},\label{char2} 
\end{eqnarray} 
such that the characteristic $(R(s,t,r,w,F),W(s,t,r,w,F),F)$ goes 
through the point $(r,w,F)$ when $s=t$. 
This representation shows that $f$ is nonnegative for all $t\geq 0,$ $\|f\|_{\infty}=\|f_{0}\|_{\infty},$ and that 
for regular initial data with angular momentum cut-off, 
$f(t,r,w,F)=0$ if $F\leq F_{-}.$ 
There are two known conserved quantities for the Einstein-Vlasov
system, conservation of the number of particles 
\begin{displaymath} 
4\pi^{2}\int_{0}^{\infty}e^{\lambda(t,r)}
\left(\int_{-\infty}^{\infty}\int_{0}^{\infty}f(t,r,w,F)dFdw\right) dr, 
\end{displaymath} 
and conservation of the ADM mass 
\begin{equation} 
M:=4\pi\int_{0}^{\infty}r^{2}\rho(t,r)dr.\label{adm} 
\end{equation} 
These conservation laws follow from general argments but can easily be
obtained by taking the time derivative of the integral expressions and
make use of the Einstein equations and the Vlasov equation. 
The mass function $m$ is defined by 
\begin{equation} 
m(t,r):=4\pi\int_{0}^{r}\eta^{2}\rho(t,\eta)d\eta,\label{m} 
\end{equation}
and by integrating (\ref{ee2}) we find 
\begin{equation} 
e^{-2\lambda(t,r)}=1-\frac{2m(t,r)}{r}.\label{e2-lambda} 
\end{equation} 
Thus, as long as the solution exists we have
\begin{equation}
\frac{m(t,r)}{r}<1/2.\label{moverrlessthan12}
\end{equation}
A fact that will be used frequently is that $$\mu+\lambda\leq 0.$$ 
This is easily seen by adding the equations (\ref{ee1}) and
(\ref{ee2}), which gives $$\lambda_{r}+\mu_{r}\geq 0,$$ and then using
the boundary conditions on $\lambda$ and $\mu.$ From (\ref{e2-lambda})
we see that $\lambda\geq 0,$ and it follows that $\mu\leq 0.$
Finally, we note that in~\cite{RR1} a local existence theorem is proved
and it will be used below that a classical (or regular) solution 
exists on some time interval $[0,T[.$ 
\section{General a priori bounds} 
In this section we show a few a priori bounds which are
easy to derive but cannot be found in the literature. Together with
the conserved quantities mentioned in section 2, and the inequality 
(\ref{moverrlessthan12}), these results constitute the only known a
priori bounds for the spherically symmetric Einstein-Vlasov system. 
\begin{lemma}
Let $(f,\mu,\lambda)$ be a regular solution to the Einstein-Vlasov system. 
Then 
\begin{eqnarray}
&\displaystyle\int_{0}^{\infty}4\pi(\rho+p)re^{2\lambda}e^{\mu+\lambda}dr\leq 1,&\label{expr}\\ 
&\displaystyle\int_{0}^{\infty}(\frac{m}{r^{2}}+4\pi p)re^{2\lambda}e^{\mu}dr\leq
1.&\label{expr2} 
\end{eqnarray} 
\end{lemma}
\textbf{Proof. }Using the boundary conditions (\ref{bdry}) and that
$\mu+\lambda\leq 0$ we get 
\begin{eqnarray*}
1\geq 
1-e^{\mu+\lambda}(t,0)&=&\int_{0}^{\infty}\frac{d}{dr}e^{\mu+\lambda}dr\\
&=&\int_{0}^{\infty}(\mu_{r}+\lambda_{r})e^{\mu+\lambda}dr. 
\end{eqnarray*}
The right hand side equals (\ref{expr}) by equations (\ref{ee1}) 
and (\ref{ee2}) which completes the first part of the lemma. The
second part follows by studying $e^{\mu}$ instead of
$e^{\mu+\lambda}.$ 
\begin{flushright}
$\Box$ 
\end{flushright} 
Next we show that not only $\rho(t,\cdot)\in L^{1},$ which follows from
the conservation of the ADM mass, but that also $e^{2\lambda}\rho(t,\cdot)\in
L^{1}.$ 
\begin{lemma}
Let $(f,\mu,\lambda)$ be a regular solution to the Einstein-Vlasov
system. 
Then 
\begin{equation}
\int_{0}^{\infty}r^{2}e^{2\lambda}\rho(t,r) dr\leq
\int_{0}^{\infty}r^{2}e^{2\lambda}\rho(0,r) dr+\frac{t}{8\pi}.\label{r2rhoe2lambda} 
\end{equation} 
\end{lemma}
\textbf{Proof. }Using the Vlasov equation we obtain 
\begin{eqnarray*} 
\partial_{t}\left(r^{2}e^{2\lambda}\rho(t,r)\right)&=& 
-\partial_{r}\left(r^{2}e^{\mu+\lambda}j\right)-re^{\mu+\lambda}2je^{2\lambda}
\frac{m}{r}\\
&\leq&-\partial_{r}\left(r^{2}e^{\mu+\lambda}j\right)+ 
\frac{1}{2}re^{\mu+\lambda}(\rho+p)e^{2\lambda}. 
\end{eqnarray*} 
Here we used that $m/r\leq 1/2$ together with the elementary inequality 
$2|j|\leq\rho+p,$ which follows from the expressions 
(\ref{rho})-(\ref{j}). In view of (\ref{char1}) we see that
$\lim_{r\to\infty}r^2j(t,r)=0,$ since the initial data has compact
support. Since the solution is regular and hence bounded the boundary
term at $r=0$ also vanishes (as a matter of fact spherical symmetry
even implies that $j(t,0)=0$). Thus, by lemma 1 we get 
\begin{displaymath}
\partial_{t}\int_{0}^{\infty}r^{2}e^{2\lambda}\rho(t,r) dr\leq 
\frac{1}{8\pi}, 
\end{displaymath} 
which completes the proof of the lemma. 
\begin{flushright}
$\Box$ 
\end{flushright} 
A natural question is if the inequalies (\ref{expr}) and
(\ref{expr2}) can bound any matter quantity in an $L^{p}-$space with
$p>1,$ in view of the fact that merely $r$ and not $r^{2}$ is present
in the integrand. For the Vlasov-Poisson- and the Vlasov-Maxwell
system such $L^{p}$ estimates (with $p>1$) play an essential role in
the global existence proofs (although the problem in $1+3$ dimensions
is open for the Vlasov-Maxwell system affirmative results exist in 
lower dimensions~\cite{GSc2}) and such results are thus desirable. We
have the following result (where
$L^{p}(R^3,\omega)$ means the $L^p$ space with weight $\omega$): 
\begin{lemma} 
Let $(f,\mu,\lambda)$ be a regular solution to the Einstein-Vlasov
system for initial data with angular momentum cut-off. Then 
$$q(t,\cdot)\in
L^{5/4}(R^{3},e^{\mu+\lambda}e^{2\lambda}).$$ 
\end{lemma} 
\textit{Proof. }Here we use the momentum variable $v$ instead of
$(w,F).$ These are related as follows
\[ 
w=\frac{x\cdot v}{r},\; F=x^{2}v^{2}-(x\cdot v)^{2}=|x\times v|^{2}, 
\]
where $r=|x|.$ From~\cite{R} it follows that $q$ now takes the form 
$$q=\int_{R^3}\frac{Ff}{r^2\sqrt{1+v^2}}dv.$$ 
By the assumption on the initial data we have that
$r\sqrt{1+v^2}\geq\sqrt{F_{-}}$ on the support of $f$ and we get 
\begin{eqnarray}
q&=&\int_{R^3}\frac{Ff}{r^2\sqrt{1+v^2}}dv\\
&\leq& \int_{|v|\leq K}\frac{Ff}{r^2\sqrt{1+v^2}}dv+\int_{|v|\geq
  K}\frac{Ff}{r^2\sqrt{1+v^2}}dv\\
&\leq& \int_{|v|\leq
  K}\frac{Ff\sqrt{1+v^2}}{r^2(1+v^2)}dv+\frac{1}{K}\int_{|v|\geq 
  K}\frac{Ff}{r^2}dv\\
&\leq& C(F_{-},F_{+},\|f\|_{\infty})K^{4}+\frac{C}{K}\int_{R^3}\frac{Ff\sqrt{1+v^2}}{r}dv\\
&\leq&C\left(\int_{R^3}\frac{Ff\sqrt{1+v^2}}{r}dv\right)^{4/5}. 
\end{eqnarray} 
Here we took $$K=\left(\int_{R^3}\frac{Ff\sqrt{1+v^2}}{r}dv\right)^{1/5}.$$ 
Now since $F\leq F_{+}$ we get that 
\begin{equation}
q^{5/4}\leq C\int_{R^3}\frac{f\sqrt{1+v^2}}{r}dv\leq C\frac{\rho}{r}, 
\end{equation} 
and from (\ref{expr}) we obtain 
\begin{eqnarray} 
&\displaystyle\int_{0}^{\infty}r^2q^{5/4}e^{\mu+\lambda}e^{2\lambda}dr\leq 
C\int_{0}^{\infty}\rho re^{2\lambda}e^{\mu+\lambda}dr\leq C,&\nonumber 
\end{eqnarray}
and the lemma follows. 
\begin{flushright} 
$\Box$
\end{flushright} 
\textit{Remark. }The first factor $e^{\mu+\lambda}$ in the weight
$e^{\mu+\lambda}e^{2\lambda}$ is the square root of the determinant of
the metric and can thus be viewed as a part of the integration
measure. The second factor $e^{2\lambda}$ is always greater than one 
which then sharpens the bound for $q,$ in analogy with lemma 2. 
%We now choose a time interval $[t_{1},t_{2}]$ such that 
%$$\frac{m(s,R(s))}{R(s)}\leq \frac{1}{3},$$ 
%for all $s\in [t_{1},t_{2}]$ with strict inequality on the open
%interval $(t_{1},t_{2}).$ 
%Consider a trajectory $(R(s),W(s),F)$ such that
%$f(t_{1},R(t_{1}),W(t_{1}),F)
%\neq 0,$ and such that $W(s)\leq 0$ on $(t_{1},t_{2}).$ 
%We have 
%\begin{eqnarray*} 
%\frac{d}{ds}R^{-1}(s)&=& 
%\left[ \frac{e^{(\mu-\lambda)(s,R(s))}}{R(s)}\frac{|W(s)|}{\sqrt{1+W^{2}+F/R^{2}}}\right]
%R^{-1}(s),\\ 
%&\leq &\frac{1}{\sqrt{F}}e^{(\mu-\lambda)(s,R(s))}|W(s)|R^{-1}(s)\\ 
%&\leq &\frac{1}{\sqrt{F_{-}}}e^{(\mu-\lambda)(s,R(s))}E(R,W,F)R^{-1}(s). 
%\end{eqnarray*} 
%
%\textbf{Remark. }In kinetic theory we have angular momentum also in the spherically
%symmetric case which is an essential difference to fluid models. 
%The importance of having angular momentum ($F>0$) is clearly seen in
%the inequality above.\\ 
\section{A comparison between the EV- and the VP system} 
It is instructive to compare the structure of the spherically
symmetric Vlasov-Poisson system, where the gravitational field is
Newtonian, with the spherically symmetric Einstein-Vlasov system
in Schwarzshild coordinates (in other coordinates such a 
comparison can be slightly different). 
The former system is the limit of the latter as the speed of light
goes to infinity, as was proved in~\cite{RR2}, but the mathematical
understanding of these two systems are nevertheless very different. In
this section we try to point out the main differencies between these
systems from a mathematical point of view. 
A Newtonian gravitational system is modelled in
kinetic theory by the Vlasov-Poisson system which reads 
\begin{eqnarray}
&\partial_{t}f+p\cdot\nabla_{x}f-E(t,x)
\cdot\nabla_{p}f=0,&\label{v}\\
&E=\nabla\phi,\;\;\;\Delta\phi=\rho:=\int_{R^3}f(t,x,p)dp.&\label{poisson}
\end{eqnarray}
In the case of spherical symmetry Batt~\cite{B} gave a global
existence proof in 1977, but later on this was also obtained for general
data, independently by Pfaffelmoser~\cite{Pf}
and Lions and Perthame~\cite{LP}. The method of proof
by Pfaffelmoser (see also~\cite{Sc2} and~\cite{Ho2}) is based on a study
of the characteristic system 
associated with the Vlasov equation. The quantity that plays a 
major role is $Q(t),$ which was defined in the introduction. 
If $Q(t)$ can be bounded by a 
continuous function in $t,$ then global existence follows. 
In~\cite{Ho2} the sharpest bound of $Q(t)$ has been given for
the system (\ref{v})-(\ref{poisson}), where it is proved that
$Q(t)$ grows at most as $C(1+t)\log{(C+t)}.$ As have already been
pointed out, the continuation criterion
in~\cite{RR1} shows that a bound of $Q(t)$ is sufficient for
global existence also in the case of the spherically symmetric
Einstein-Vlasov system. It is therefore natural to compare the
characteristic systems in the two cases, since the natural way to
control $Q(t)$ is to obtain bounds on the solutions to this 
system. In the case of the spherically symmetric Einstein-Vlasov
system this means to bound $E(t)=\sqrt{1+W^{2}(t)+F/R^{2}(t)}.$ 
It follows 
from the characteristic equations (\ref{char1}) and (\ref{char2}) 
that $E$ satisfies 
\begin{equation}
\frac{d}{ds}E(s)=(-W)\left[e^{\mu+\lambda}\left(\frac{m}{R^{2}}
+ 4\pi R
  \left(\frac{(-W)}{E}j+p\right)\right)\right].\nonumber 
\end{equation} 
Here 
\[ m(t,r)=\int_{0}^{r}\eta^2\rho(t,\eta)\,d\eta. \] 
A comparison with the analog equation for the Vlasov-Poisson system
leads to the following observations: 
\begin{itemize}
\item the factor (-W) is only present in EV not VP 
\item only the m-term is present in VP 
\item the dependence on $j$ and $p$ is pointwise 
\end{itemize}
Having a Gr\"{o}nvall estimate in mind the first point is a severe
difficulty for the Einstein-Vlasov system since the magnitude of $|W|$ 
and $E$ might be of the same order, and thus gives much less room for
estimating the right-hand side. The second and third point concern
the regularity of the right hand side. The quasi local mass $m$ is an
average of the energy density $\rho$ and is thus more regular than
$\rho$ itself, and this fact is important in both the Pfaffelmoser- 
and the Lions-Perthame methods of proof. In the EV
case there is a pointwise dependence on the matter terms $j$ and $p$
which indeed is a difficulty but as will be seen in section 5 this
dependence is in a sense mild. 
Other additional difficulties are the following facts: 
\begin{itemize}
\item $Q^{4}$ instead of $Q^3$ bound the matter terms $\rho,|j|,$ and $p$
\item conservation of ADM mass does not lead to an $L^p$ bound of
  $\rho$ with $p>1$ 
\end{itemize}
The densities in the two cases are (in Minkowski space) given by 
$$\rho^{VP}=\int_{R^3}fdp,\mbox{ and }\rho^{EV}=\int_{R^3}\sqrt{1+|p|^2}fdp.$$ 
Again, having a Gr\"{o}nvall estimate in mind it is a disadvantage that
$\rho^{EV}$ (and the other matter terms) are bounded in terms of $Q^4$
instead of $Q^3$ as is the case for $\rho^{VP}.$ Conservation of
energy in the VP case means that $$\int_{R^3}\int_{R^3}|p|^{2}fdpdx,$$
is bounded 
and this in turn implies that $\rho^{VP}(t,\cdot)\in L^{5/3},$ whereas
conservation of the ADM mass does not qualify $\rho^{EV}$ into any $L^{p}$
space with $p>1.$ In both methods of proof in the VP case such an 
a priori bound plays a fundamental role. 
% i radial + maximal gauge så lägg till resultat för RVP att
% \rho^{RVP}\in L^{3/2+} är tillräckligt. 
We conclude this section by a comparison with the relativistic
Vlasov-Poisson system. 
By starting from the relativistic Vlasov-Maxwell system, which describes a
charged plasma, and assuming spherical symmetry the following system
is obtained 
\begin{eqnarray}
&\partial_{t}f+\hat{p}\cdot\nabla_{x}f+\beta E(t,x)
\cdot\nabla_{p}f=0,&\label{rv}\\
&E=\nabla\phi,\;\;\;\Delta\phi=\rho:=\int_{R^3}f(t,x,p)dp.&\label{rpoisson}
\end{eqnarray}
Here $\hat{p}=p/\sqrt{1+p^2}$ and $\beta=1.$ This system makes sense also
without the assumption of spherical symmetry as well as with 
$\beta=-1$ and is called the relativistic Vlasov-Poisson system with a
repulsive ($\beta=1$) or attractive ($\beta=-1$) potential. This
system is not Lorentz invariant and is not a physically correct
equation (see~\cite{An3} for a discussion of this point) but from a
formalistic point of view \textit{the only difference} from the VP 
system is that the $p$ in front
of the $\nabla_{x}f-$term has been changed to $\hat{p},$ and in this
sense resembles the EV system. Nevertheless,
the mathematical results in the two cases are very different. 
The most relevant results in this context can be summarized as follows
(we refer to~\cite{GSc1},~\cite{GSc4} for details): 
\begin{itemize}
\item $\beta=1.$ A general global existence result is still
missing for the RVP system but the spherically- and the cylindrically
symmetric cases have been settled. 
\item $\beta=-1.$ Solutions blow up in finite time for a large class
  of initial data. 
\end{itemize}
We see that even in the repulsive case much less is known than 
for the Vlasov-Poisson system, and apparently, in the more relevant attractive
case, the situation is very different. 
In conclusion, it is far from clear that the well-developed machinery
for the classical Vlasov-Poisson system can contribute much to the
understanding of the spherically symmetric Einstein-Vlasov system. 
Furthermore, the blow-up result for the relativistic Vlasov-Poisson
system with $\beta=-1$ indicates that if the Einstein-Vlasov system
prevents blow-up in finite time, the method for proving it might be
quite subtle. 
%\section{Global existence if $m/r\leq 1/3$ and $j\leq 0.$} 
% kan gå även med de termer som innefattar W/E-bidraget ty \lambda
% begränsad ger att d/dt av r\rhoe^{2\lambda} kan betraktas och vi har
% då termer av typen \lambda_{t}\lambda_{r} som är OK. Noter att min
% uträkning ger precis att den ouppklarade termen är \lt\lr. 
% kräver dock ingoing ty j används i uppskattningen 
% ... m/r\geq 1/3 överallt kräver j\leq 0 
% argumenten i T_{2} är olika varför detta ändå inte funkar 

%\section{Oscillationer, ursprungsmetod} 
% i) anta \mu, \l begränsade och ingoing, då är det klart. 
% ii) anta ingoing och ändligt antal oscillationer. 
% iii) arb: anta \mu,\l <C, visa att detta är tillräckligt. Notera att
% om vi omvandlar j+p termen till domän så ger e^\muE en -p-term som
% kills off k längs karaktäristiska. Ursprungsmetod. 

\section{A mixed $L^{3+}-$ and weighted $L^{\infty}$ approach} 
% ange också geometrisk betydelse av huvudterm. När vi väljer mellan
% inre och yttre område får vi olika för och nackdelar. Inre: OM vi
% antar bara W<0 så räcker estimat på q, vi får bra randvärden om
% trajektorian når längst. Yttre: om vi tar en trajektoria som vänder
% så får vi bra randvärde ty W=0 och med (L^p)-antagande får vi
% universal kontroll på vändpunkt och därmed på W^{+}. Detta gör att
% vi kan klara randvärdet i allmänhet och kan därmed behandla
% godtycklig trajektoria och vi behöver ``bara'' L^3 ej
% +epsilon. Däremot behövs estimat på rho istället för bara q. 
In this section we make assumptions on the matter quantities and show
that these are sufficient for bounding $Q(t).$ Since these assumptions
are satisfied when $Q(t)$ is bounded we obtain an improved
continuation criterion. From the proof it will be clear that
the pointwise dependence on $j$ and $p$ in the characteristic
equation, discussed in the previous section, is in a sense mild,
i.e. as long as $j$ and $p$ do not contain too sharp and narrow peaks
then their contributions can be controlled. 

Assume that there exist constants $A$ and $B$ such that for $t\in [0,T[,$ 
\begin{eqnarray} 
%% &\lambda(t,\cdot)\in L^{\infty}(R^{3}),&\label{hyplambdalessC}\\ 
&r^2k(t,r)\leq A,&\label{hypsupr2}\\ 
&\|\rho(t,\cdot)\|_{L^{3+}(B_{T}^{3})}\leq B.&\label{hyprhoLp}
\end{eqnarray} 
%% $\lambda$ begränsat följer av de andra två 
Here $B_{T}^{3}$ is the ball $\{x:|x|\leq R_{+}+T\}$ 
and $L^{3+}$ means that it is sufficient that $\rho\in 
L^{3+\epsilon}$ for any given $\epsilon>0.$ Recall also the definition
of $k$ in equation (\ref{k}). 
First note that these assumptions are weaker than the assumption that
$Q(t)$ is bounded on $[0,T[.$ Indeed, from (\ref{rho}) it follows that 
$\rho\leq CQ(t)^4.$ 
Therefore a bound on $Q(t)$ implies that 
\begin{equation}
\rho(t,r)\leq C, \mbox{ on }[0,T[,\label{rhoboundedbyC} 
\end{equation} 
which in view of the fact that the severe difficulties are
concentrated at $r=0,$ is significantly stronger than (\ref{hypsupr2})
(recall $k\leq\rho$). 
That (\ref{hyprhoLp}) follows from (\ref{rhoboundedbyC}) is immediate. 
It is also instructive to see that (\ref{hypsupr2}) is a natural
hypothesis in view of the fact that 
\begin{equation} 
\frac{m}{r}\leq\frac{1}{2} \mbox{ on }[0,T[.\label{moverr} 
\end{equation}
From the boundary condition $\lambda(t,0)=0$ it also follows that 
$$\frac{m}{r}\to 0, \mbox{ as }r\to 0.$$ 
If we assume that there exists a constant $C$ and a positive number
$\alpha$ such that $$\rho(t,r)\leq\frac{C}{r^{\alpha}},$$ and then
estimate $m/r$ we get 
\begin{equation}
\frac{m}{r}=\frac{\int_{0}^{r}\eta^2\rho(t,\eta)d\eta}{r}\leq Cr^{2-\alpha}.
\end{equation}
Thus, if such an estimate for $\rho$ exists then $\alpha\leq 2,$ which 
shows that (\ref{hypsupr2}) is natural. However, $j$ or $p$ might have 
sharp, and then by (\ref{moverr}) necessarily narrow, peaks so that
such an estimate is not valid at all. We have not been able to exlude
that such peaks exist which is the reason for imposing assumption
(\ref{hypsupr2}). This assumption still allows for sharp peaks in the
matter quantity $q,$ which is the part of $\rho$ which is associated
to the tangential momenta. The potential difficulty of sharp peaks in $j$
and $p$ is of major interest to understand and it might be that this
difficulty is a particular feature of the Einstein-Vlasov system in
Schwarzschild coordinates. We have the following theorem. 
\begin{theorem}Consider a solution of the Einstein-Vlasov system for
  regular initial data with radial cut-off on its maximal time
  interval $[0,T[$ of existence. Assume that (\ref{hypsupr2}) and
  (\ref{hyprhoLp}) hold. Then $Q(t)$ is bounded on $[0,T[,$ and
  $T=\infty.$ 
\end{theorem} 
%\textbf{Remark. }Here $q\in L^{3+}$ means that it is sufficient that
%$q\in L^{3+\epsilon}$ for any $\epsilon>0.$ 
\textit{Proof. }
The quantity that will be considered is $E,$ and the equation along a 
characteristic $(R(t),W(t),F)$ for $E$ reads 
\begin{equation}
\frac{d}{ds}E(R(s),W(s),F)=\frac{(-W)}{E}\left[e^{\mu+\lambda}\frac{m}{R^{2}}
+ 4\pi Re^{\mu+\lambda}\left(
  \frac{(-W)}{E}j+p\right)\right]E.\label{charE} 
\end{equation} 
Note that we often choose to write $(-W)$ since the major case of
interest is when the characteristic is ingoing, i.e. $W<0.$ 
It follows that 
\begin{equation} 
\log{E(T)}=\log{E(0)}+
\int_{0}^{T}\frac{(-W)(t)}{E(t)}\left[e^{\mu+\lambda}\frac{m}{R^{2}}
+ 4\pi Re^{\mu+\lambda}\left(
  \frac{(-W)}{E}j+p\right)\right]\, dt.\label{ET} 
\label{curvein} 
\end{equation} 
The maximal time interval of existence is $[0,T[$ and we can assume
that $T<\infty.$ Then there exists a characteristic $(R(t),W(t),F)$ and a 
$T_{0}\in [R_{-},T]$ such that $$\lim_{t\to T_{0}}R(t)=0,$$ since
otherwise the finiteness of $T$ would contradict the result
in~\cite{RRS1}. (It is not necessary for our arguments that the limit
do exist when $T_{0}=T,$ but due to the causal character of spacetime
it exists.) 
We consider the set of all characteristics with this property and we pick a
characteristic $(R^{*}(t),W^{*}(t),F^{*})$ in this 
set with an associated time $T^{*}_{0}$ such that there
is no other characteristic in this set with an associated $T_{0}$ strictly
less than $T^{*}_{0}.$ Hence we have the following properties of
$(R^{*}(t),W^{*}(t),F^{*})$ and $T^{*}_{0},$ 
\begin{equation} 
\lim_{t\to T^{*}_{0}}R^{*}(t)=0,\label{limT0} 
\end{equation} 
and
\begin{equation}
f(t,0,w,F)=0 \mbox{ for all } t<T_{0}^{*}.\label{emptyaxis} 
\end{equation} 

From now on we drop the superscript $*$ and consider a characteristic 
$(R(t),W(t),F)$ with the properties above. 
From the equations (\ref{ee2}) and (\ref{e2-lambda}) we have 
\begin{equation} 
\mu_{r}(t,r)=(\frac{m}{r^2}+4\pi rp)e^{2\lambda},\label{mur} 
\end{equation}
and by using (\ref{ee3}) the time integral in (\ref{ET}) can be written 
\begin{equation} 
\int_{0}^{T}\frac{(-W(t))}{E(t)}\left[e^{\mu-\lambda}\mu_{r}(t,R(t))
-\lambda_{t}(t,R(t))\frac{(-W(t))}{E(t)}\right]\, dt.\label{t-in} 
\end{equation}
Let us denote the curve $(t,R(t)),\; 0\leq t\leq T_{0},$ by $\gamma.$
By using that $$\frac{dR}{dt}=-e^{\mu-\lambda}\frac{(-W)}{E},$$ the
integral above can be written as a curve integral 
\begin{eqnarray} 
&\displaystyle \int_{\gamma}\frac{(-W(t))}{E(t)}e^{(-\mu+\lambda)(t,r)}\lambda_{t}(t,r)dr+
\frac{(-W(t))}{E(t)}e^{(\mu-\lambda)(t,r)}\mu_{r}(t,r)dt.&\label{curve1} 
\end{eqnarray} 
Let $C$ denote the closed curve 
$\gamma+C_{2}+C_{3},$ oriented counter clockwise, 
where $$C_{2}=\{(t,0):T_{0}\geq t\geq 0\},$$ and
$$C_{3}=\{(0,r):0\leq r\leq R(0)\}.$$ Let $\Omega$ denote the
domain enclosed by $C.$ 
We get by Green's formula in the plane 
\begin{eqnarray} 
&\displaystyle\oint_{C}\frac{(-W(t))}{E(t)}e^{-\mu+\lambda}\lambda_{t}dr+
\frac{(-W(t))}{E(t)}e^{\mu-\lambda}\mu_{r}dt&\nonumber\\ 
&\displaystyle =\int\int_{\Omega}\frac{(-W(t))}{E(t)}\left[\partial_{r}
\left(e^{\mu-\lambda}\mu_{r}-
\partial_{t}\left(e^{-\mu+\lambda}
\lambda_{t}\right)\right)\right]dtdr&\nonumber\\ 
&\displaystyle
-\int\int_{\Omega}\frac{d}{dt}\left(\frac{(-W(t))}{E(t)}\right)
\left(e^{-\mu+\lambda}\lambda_{t}\right)\,dtdr.&\label{green} 
\end{eqnarray} 
Note that the factor $W(t)/E(t),$ which really only is defined along
the characteristic, has been considered as a function of
$t$ alone, i.e. for given $t$ it is defined to be constant in $r$ with
the value it has at $(t,R(t)).$ 
Alternatively, $E$ could have been viewed as a function of $t$ and
$r,$ by taking $E=\sqrt{1+W^{2}(t)+F/r^2}.$ The final conclusion would
of course be the same. 
An interesting fact is now that the first term in the right hand side
which contains a combination of second order derivatives of $\mu$ and
$\lambda$ can be substituted by one of the Einstein equations
(\ref{ee4}). Indeed, we get 
\begin{eqnarray} 
&\displaystyle\int\int_{\Omega}\frac{(-W(t))}{E(t)}\left(\partial_{r}
\left( e^{\mu-\lambda}\mu_{r}\right)-\partial_{t}\left(e^{-\mu+\lambda}
\lambda_{t}\right)\right)dtdr&\nonumber\\ 
&\displaystyle
=\int\int_{\Omega}\frac{(-W(t))}{E(t)}\left((\mu_{rr}+(\mu_{r}-\lambda_{r}))e^{\mu-\lambda}-\lambda_{tt}-(\mu_{t}-\lambda_{t})\right)e^{-\mu+\lambda}\,dtdr& 
\nonumber\\
&\displaystyle =\int\int_{\Omega}\frac{(-W(t))}{E(t)}\left(4\pi
qe^{\mu+\lambda}-(\mu_{r}-\lambda_{r})\frac{e^{\mu-\lambda}}{r}\right)\,dtdr&
\nonumber\\ 
&\displaystyle =\int\int_{\Omega}\frac{(-W(t))}{E(t)}e^{\mu+\lambda}
\left(4\pi(\rho-p)+4\pi q-\frac{2m}{r^{3}}\right)\,dtdr.&\label{id}
\end{eqnarray} 
Here we used (\ref{mur}) for $\mu_{r}$ and that 
$$\lambda_{r}=(4\pi\rho-\frac{m}{r^2})e^{2\lambda},$$ which follows
from (\ref{ee1}). 
By using the characteristic equations (\ref{char1}) and (\ref{char2})
we get that the second term in (\ref{green}) is given by 
\begin{equation} 
-\int\int_{\Omega}\left(\left(\mu_{r}e^{\mu-\lambda}+\lambda_{t}\frac{(-W(t))}{E(t)}\right)\left(1-\frac{W^2}{E^2}\right)-\frac{Fe^{\mu-\lambda}}{R^3E^2}\right)e^{-\mu+\lambda}\lambda_{t}dtdr.\label{WE-bidrag}
\end{equation} 
To summarize we have obtained the following identity 
\begin{eqnarray} 
&\displaystyle\oint_{C}\frac{(-W(t))}{E(t)}e^{-\mu+\lambda}\lambda_{t}dr+
\frac{(-W(t))}{E(t)}e^{\mu-\lambda}\mu_{r}dt&\nonumber\\
&\displaystyle =\int\int_{\Omega}\frac{(-W(t))}{E(t)}e^{\mu+\lambda}
\left(4\pi(\rho-p)+4\pi q-\frac{2m}{r^{3}}\right)\,dtdr&\nonumber\\ 
&\displaystyle
-\int\int_{\Omega}\left[\left(\mu_{r}e^{\mu-\lambda}-\lambda_{t}\frac{(-W(t))}{E(t)}\right)\left(1-\frac{W^2}{E^2}\right)-\frac{Fe^{\mu-\lambda}}{R^3E^2}\right]e^{-\mu+\lambda}\lambda_{t}\,dtdr.&\nonumber\\
&\displaystyle =:T_{1}+T_{2}.&\label{hid} 
\end{eqnarray}
%%%%%%%%%%%%%%%%%%%%%%%%%%%%%%%%%%%%%%%%%%%%%%%%%%%%
% andra termen har i sin faktor (1-w^2/E^2) precis strukturen i GMP,
% kanske kan (L2) ändras till annat antagande 
%%%%%%%%%%%%%%%%%%%%%%%%%%%%%%%%%%%%%%%%%%%%%%%%%%
Note that the function in square brackets in $T_{2}$ only depends on
$t$ and its argument is $(t,R(t)),$ 
whereas it is $(t,r)$ for $e^{-\mu+\lambda}\lambda_{t}.$ 
The term $T_{1}$ must be considered as the main term. If the integrand in 
(\ref{t-in}) is nonnegative we can drop the factor 
$(-W(t))/E(t)$ and estimate the time integral by $T_{1}$ alone (as in
section 7). More 
importantly, $T_{1}$ has a geometrical meaning; ignoring the factor
$(-W)/E$ it is the integral of the Gauss curvature of the two
dimensional manifold $R^{+}_{t}\times R^{+}_{r}$ with metric
$ds^{2}=-e^{2\mu}dt^{2}+e^{2\lambda}dr^{2}.$ (We really have in mind
the formula in (\ref{id}) obtained before we used Einstein's equations
to  get an expression in terms of matter quantities.) It is
interesting to note that the principal term in the work on BV 
solutions of the ESF system by Christodoulou~\cite{Cu3} is of the same
form, and in complete analogy with our result (see p. 1172
in~\cite{Cu3}). 

By our assumption that $\rho\in L^{3+},$ it is an easy matter to
estimate $T_{1}$ as it stands. However, we may first note that in 
the case of initial data with angular momentum cut-off, $T_{1}$ can
be simplified since the following relation holds 
\begin{displaymath} 
\rho(t,r)-p(t,r)=q(t,r)+\int_{-\infty}^{\infty}\int_{0}^{\infty}
\frac{\pi}{r^{2}E}f(t,r,w,F)dFdw, 
\end{displaymath} 
where the last term is harmless. Indeed, note that 
$$\int_{-\infty}^{\infty}\int_{0}^{\infty}
\frac{\pi}{r^{2}E}f(t,r,w,F)dFdw\leq\frac{1}{F_{-}}r^2\rho,$$ where we
used the property that $F\geq F_{-}.$ 
This term is bounded in terms of the ADM mass. 
Therefore, the remaining and the principal part, $T_{1}^{prin},$ of
$T_{1}$ reads 
\begin{displaymath}
T_{1}^{prin}:=\int\int_{\Omega}\frac{(-W(t))}{E(t)}e^{\mu+\lambda}
\left(8\pi q-\frac{2m}{r^{3}}\right)\,dtdr, 
\end{displaymath} 
which has the advantage that only $q$ and not $\rho$ is explicitly in
the integrand ($m$ depends of course on $\rho$). This will have
significance in section 7. 

We now estimate $T_{1}$ in (\ref{hid}). By using $|W|/E<1$ and
$e^{\mu+\lambda}\leq 1,$ we have 
\begin{eqnarray}
|T_{1}|&\leq&\int\int_{\Omega}
\left(\frac{2m}{r^{3}}+8\pi \rho\right)\,dtdr=\int_{0}^{T}\int_{0}^{R(t)}\left(\frac{2m}{r^{3}}+8\pi
\rho\right)\,drdt. 
\nonumber\\ 
\end{eqnarray} 
It should be noted that it is our assumption (\ref{hyprhoLp}) on
$\rho$ which allows for this rough estimate. In a more 
careful treatment the major difficulty should
be associated with an ingoing trajectoria, i.e. $W<0,$ and
therefore the signs of the terms in the integrand should be important 
(see section 7). 
By our assumption (\ref{hyprhoLp}) we get by H\"{o}lder's inequality
($1/p+1/p'=1,\; p>3$) 
\begin{eqnarray} 
&\displaystyle m=\int_{0}^{r}\eta^{2}\rho(t,\eta)\,d\eta\leq
\left(\int_{0}^{r}\eta^{2}d\eta\right)^{1/p'}\left(\int_{0}^{r}\eta^{2}\rho^{p}d\eta\right)^{1/p}&\nonumber\\
&\displaystyle\leq
Cr^{2+}\left(\int_{0}^{r}\eta^{2}\rho^{p}d\eta\right)
^{1/p},& 
\end{eqnarray}
and 
\begin{eqnarray}
&\displaystyle\int_{0}^{R(t)}\rho(t,r)\,dr\leq\left(\int_{0}^{R(t)}r^{-2p'/p}dr\right)^{1/p'}\left(\int_{0}^{R(t)}r^{2}\rho^{p}dr\right)^{1/p}&\nonumber\\
&\displaystyle\leq
C\left(\int_{0}^{R(t)}r^{2}\rho^{p}dr\right)^{1/p},&\label{qestimate} 
\end{eqnarray} 
where we have used that $p'/p<1/2.$ 
In view of these estimates we obtain 
\begin{equation}
|T_{1}|\leq CBT. 
\end{equation} 
Now we consider the term $T_{2}.$ We begin with a lemma showing that
our hypotheses imply that $\lambda$ is bounded. 

%
% Lemmat är OK med Lp, p>3/2, och [RRS] ty j\leq g(R). 
%

%
% Lyft ut lemmat och kanske ha conjecture om p>3/2 samt kommentera att
% RVP ok om p\geq 3/2. 
% 

\begin{lemma}Assume that (\ref{hypsupr2}) and (\ref{hyprhoLp}) hold on 
$[0,T[.$ Then $\lambda$ is uniformly bounded on $[0,T[.$ 
\end{lemma}
\textit{Proof. }From the relation (\ref{e2-lambda}) we see that
$\lambda$ is bounded as long as $m/r$ is strictly less than $1/2.$
From (\ref{hyprhoLp}) it follows that 
\[ 
\frac{m}{r}\leq\frac{r}{3^{2/3}}\|\rho\|_{L^{3}}\leq\frac{rB}{3^{2/3}}.
\] 
If $r\leq 3^{2/3}/4B$ it follows that $m/r\leq 1/4$ and
$\lambda$ is bounded by $\log{2}/2.$ For $r\geq 3^{2/3}/4B$ we
consider equation (\ref{ee3}) and in view of (\ref{hypsupr2}) we get 
\[ 
|\lambda_{t}|\leq\frac{4\pi A}{r}\leq\frac{16\pi AB}{3^{2/3}}, 
\] 
and since $\lambda$ is bounded initially it follows that $\lambda$ is
bounded on $[0,T[$ when $r\geq 3^{2/3}/4B.$ 
By combining these estimates the lemma follows. 
\begin{flushright}
$\Box$
\end{flushright}
To estimate $T_{2},$ note again that the argument of the part
of the integrand in square brackets is $(t,R(t))$ while the argument of
$e^{-\mu+\lambda}\lambda_{t}$ is $(t,r).$ By using
$$\mu_{r}(t,r)=(m/r^2+4\pi 
rp)e^{2\lambda},$$ 
we obtain 
four terms in the integrand of $T_{2}.$ Since 
$$\lambda_{t}(t,r)=-4\pi re^{\mu+\lambda}j,$$ and the sign of $j$ is
indefinite we cannot drop any of these terms. For the first term we
have 
\begin{eqnarray}
&\displaystyle\left|\int\int_{\Omega}\left[\frac{m(t,R(t))}{R(t)^2}e^{(\mu+\lambda)(t,R(t))}\right]e^{(-\mu+\lambda)(t,r)}
e^{(\mu+\lambda)(t,r)}4\pi r(-j(t,r))\,dtdr\right|&\nonumber\\
&\displaystyle\leq
\int\int_{\Omega}\left[\frac{m}{R(t)}e^{\mu+\lambda}\right]e^{2\lambda}4\pi
|j|\,dtdr.& 
\end{eqnarray} 
Here we used that $r\leq R(t).$ 
The expression in square brackets is bounded by $1/2$ and since $\lambda$ is
bounded in view of Lemma 5, this term is bounded by 
$$C\int\int_{\Omega}|j|drdt\leq C\|\rho\|_{L^{3+}}T\leq CBT.$$ 
The remaining terms are estimated in a similar fashion by again using that
$r\leq R(t)$ together with (\ref{hypsupr2}) to estimate the terms 
$rR(t)p(t,R(t)),$ and $rR(t)|j|.$ 
In conclusion we have obtained that $$T_{1}+T_{2}\leq CBT.$$ 
Therefore, a bound on $E(t)$ follows from (\ref{ET}) if we can bound 
the curve integrals over $C_{2}$ and $C_{3}.$ The curve integral over
$C_{3}$ is obviously bounded in terms of the initial data and the curve
integral over $C_{2}$ is zero by (\ref{emptyaxis}). This
completes the proof of Theorem 1. 
\begin{flushright}
$\Box$
\end{flushright} 
\textit{Remark. }We have seen that our assumption on $\rho$ makes the
estimate easy and we have only to consider the $r-$integration in the
$T_{1}$ term. It is however clear that a more careful treatment must
benefit from the time integration. Indeed, if $\rho$ behaves like
$r^{-3/2},$ then an approach that only focus on the $r-$integration
in the $T_{1}$ term will fail while such a behaviour of $\rho$ is
perfectly allowed by (\ref{moverrlessthan12}) and the a priori bound
in Lemma 1. The methods of proof of global existence in the case of
the Vlasov-Poisson system make use of such a time integration in a
cruscial way. 
% kanske måste vi ändra här: givet epsilon>0 så finns T_{epsilon}<T
% och trajektoria (R,W) med R(T_{epsilon})\leq epsilon och sådan att
% ingen annan trajektoria varit längre in tidigare. Nu eftersom alla
% funktioner är reguljära för t\leq T_{epsilon} så kan vi göra vårt
% argument och vi får R(T_{epsilon})\geq C(T) där C är oberoende av
% epsilon och T_{epsilon}. 

\section{An approach along the characteristics} 
In this section we give conditions on $\mu_{r}$ and $\lambda_{r}$
along a characteristic which guarantee the boundedness of $Q(t).$
These conditions constitute a borderline; 
we will
show that when these conditions are satisfied, $W(t)$ must become $0$
before $R(t)=0,$ and thus the characteristic ``turns'' from ingoing 
to outgoing, and if they are not satisfied, we show under
additional hypotheses that $e^{\mu-\lambda}$ becomes sufficiently
small to make $R(t)$ stay uniformly away from zero. 
These results clearly support the idea that there is a mechanism 
which prevents blow-up of the solutions. 

Define 
\[
g(r)=\frac{1}{r}(1+\frac{1}{\log{r}}),\; 0<r<1/3. 
\] 
%and note that 
%\[
%\int g(r)=\log{(r\log{1/r})}. 
%\]
We make the following hypotheses. Along a 
characteristic $(t,R(t),W(t),F)$ we assume that the following
conditions hold when $R(t)<1/3,$ 
\begin{eqnarray} 
&\displaystyle\mu_{r}(t,R(t))\leq g(R(t))&\label{hypmur}\\ 
&\displaystyle\mu_{r}(t,R(t))-\lambda_{r}(t,R(t))\leq\frac{1}{R(t)}.&\label{hypmur-lr}
\end{eqnarray} 
The condition $R<1/3$ is only technical and we could have taken any
$R<C$ by minor changes of the presentation. 
We are going to show that $Q(t)$ is bounded under these hypotheses but
first let us discuss these assumptions and put them into a context. In
spirit these assumptions are similar to the hypothesis (\ref{hypsupr2}) in
the previous section. Observe that 
\[
\mu_{r}=(\frac{m}{r^2}+4\pi rp)e^{2\lambda}, 
\]
and we see that (\ref{hypmur}) implies
that 
\[
4\pi e^{2\lambda}p\leq\frac{g(R(t))}{R(t)}. 
\]
Thus, since $g(r)$ is roughly $1/r$ this is of the same order as
(\ref{hypsupr2}) with the main differencies that the constant $A$ is now
precisely specified and that the factor $e^{2\lambda}$ is included.
In this situation the factor $e^{2\lambda}$ is actually bounded by $3,$
since by (\ref{hypmur}) (for $R<1/3$) 
$$\frac{m}{R^2}e^{2\lambda}\leq 1/R,$$ 
and using (\ref{e2-lambda}) it follows that $m/R\leq 1/3$ so that
$e^{2\lambda}\leq 3.$ 

A natural question is what can happen if these assumptions are not
satisfied and the following lemma gives a partial answer. 
\begin{lemma}
Let $(R(t),W(t),F)$ be any solution to the characteristic system on a
time interval $I=[t_{1},t_{2}]$ with $1/3>R(t_{1})\geq\delta>0$ and 
$W(t)\leq 0$ on  $I,$ and assume either that 
\begin{displaymath}
\mu_{r}(t,R(t))\geq g(R(t))\mbox{ and }\mu_{t}(t,R(t))\leq 0, 
\end{displaymath} 
or that 
\begin{displaymath}
\mu_{r}(t,R(t))-\lambda_{r}(t,R(t))\geq\frac{1}{R(t)}\mbox{ and
}\mu_{t}(t,R(t))-\lambda_{t}(t,R(t))\leq 0, 
\end{displaymath}
on $I.$ Then there exists an $\epsilon>0,$ which only depends on
$\delta$ such that 
\[
R(s)\geq\epsilon,\; s\in[t_{1},t_{2}]. 
\]
\end{lemma}
It should be noted that by making assumptions on the signs of both 
$\mu_{t}$ and $\lambda_{t}$ along the \textit{entire} characteristic
(note that in Lemma 5 these assumptions are only made on that part
where the spatial derivatives are large) one can easily
get control of ingoing characteristics with $F>0.$ We 
formulate this in a lemma. 
\begin{lemma}Let $(R(t),W(t),F)$ be any solution to the characteristic
  system on a time interval $I=[t_{1},t_{2}]$ with 
  $R(t_{1})\geq\delta>0,\; F>0,$ and $W(t)\leq 0$ on $I.$ 
Assume that $\mu_{t}(t,R(t))\leq 0$
and that $\lambda_{t}(t,R(t))\geq 0.$ 
Then there exists an $\epsilon>0,$ which only depends on
$\delta$ such that 
\[
R(s)\geq\epsilon,\; s\in[t_{1},t_{2}]. 
\]
\end{lemma} 
Before giving the proofs of these lemmas let us discuss the sign
assumptions in Lemma 5 and compare the content of these two lemmas. 
The assumption that $\mu_{t}\leq 0$ when
$\mu_{r}\geq g$ (or $\mu_{t}-\lambda_{t}\leq 0$ when
$\mu_{r}-\lambda_{r}\geq 1/R$) is of
course the weak part of lemma 5. However, note that it always holds
that $\mu\leq 0$ (or $\mu-\lambda\leq 0$) and 
that for collapsing matter where the density is increasing, 
$\mu$ (or $\mu-\lambda$) would decrease and in this sense the case
$\mu_{t}(t,R(t))\leq 0$ (or $\mu_{t}(t,R(t))-\lambda(t,R(t))\leq 0$) 
should correspond to a regime where most matter collapse and
little disperse. At least intuitively, this should be the worst
case. Since the sign assumption in Lemma 5 is only made on the part of
the characteristic where the spatial derivative is large it
strengthens this picture since it is in the collapsing regime that
these derivatives are expected to be large. This is indeed supported by
the numerical simulation in~\cite{RRS2}. Moreover, in Lemma 6 we have
taken sign assumptions on both $\mu_{t}$ and $\lambda_{t}$ instead of
only on $\mu_{t}$ (or on $\mu_{t}-\lambda_{t}$). Lemma 6 also 
requires initial data with cut-off ($F>0$), whereas the proof of Lemma
5 holds generally. From this discussion we therefore consider Lemma 5
as being of more interest than Lemma 6. 

\textit{Proof of Lemma 5. }Let us first consider the case when $\mu_{r}\geq
g(R).$ 
Writing the characteristic equation for $R(s)$ as 
\begin{displaymath} 
\frac{dR^{-1}(s)}{ds}=\left[ \frac{e^{\mu-\lambda}}{R(s)}\frac{(-W(s))}{E(R,W,F)}\right]
R^{-1}(s), 
\end{displaymath} 
and noting that $$(-W)/E<1,$$ 
we see that $R^{-1}(s)$ can be controlled 
if we can show that 
\begin{displaymath} 
\frac{e^{(\mu-\lambda)(t,R(t))}}{R(t)}\leq C\log{\frac{1}{R(t)}}. 
\end{displaymath} 
Now 
\begin{displaymath} 
\frac{e^{(\mu-\lambda})(s,R(s))}{R(s)}\leq
\frac{e^{\mu}(s,R(s))}{R(s)}, 
\end{displaymath} 
since $\lambda\geq 0,$ 
and we have 
\begin{eqnarray} 
\frac{d}{ds}\frac{e^{\mu(s,R(s))}}{R(s)}&=&
\left(\mu_{t}+\mu_{r}\frac{dR}{ds}\right)
\frac{e^{\mu}}{R}
-\left( \frac{W\,e^{\mu-\lambda}}{E\,R}\right) \frac{e^{\mu}}{R}\nonumber\\ 
\displaystyle
&=&\left(\mu_{r}(t,R(t))-\frac{1}{R(t)}\right)
\frac{e^{\mu-\lambda}We^{\mu}}{ER}\nonumber\\
\displaystyle
&\leq&\frac{1}{R(t)\log{(1/R(t))}}(-\frac{dR}{ds})\frac{e^{\mu}}{R}.\nonumber 
\end{eqnarray} 
Here the assumptions on $\mu_r$ and $\mu_t$ were used. 
Hence we get for $t\in [t_{1},t_{2}],$ 
$$\frac{e^{\mu(t,R(t))}}{R(t)}\leq\frac{e^{\mu(t_{1},R(t_{1}))}}{R(t_{1})}\;\frac{\log{R(t)}}{\log{R(t_{1})}},$$ 
which completes the first part of the lemma. In the second case 
the proof is analogous but easier since it follows that 
$$\frac{d}{ds}\left(\frac{e^{\mu(s,R(s))-\lambda(s,R(s))}}{R(s)}\right)\leq
0,$$ 
so that $$\frac{e^{(\mu-\lambda)(t,R(t))}}{R(t)}$$ is nonincreasing on 
$[t_{1},t_{2}].$ 
\begin{flushright} 
$\Box$ 
\end{flushright} 
\textit{Proof of Lemma 6. }The spirit is the same as in the previous
proof. 
Since $F>0$ we have 
\begin{eqnarray*} 
\frac{d}{ds}R^{-1}(s)&=& 
\left[ \frac{e^{(\mu-\lambda)(s,R(s))}}{R(s)}\frac{|W(s)|}{\sqrt{1+W^{2}+F/R^{2}}}\right]
R^{-1}(s)\\ 
&\leq &\frac{1}{\sqrt{F}}e^{(\mu-\lambda)(s,R(s))}|W(s)|R^{-1}(s)\\ 
&\leq &\frac{1}{\sqrt{F}}e^{(\mu-\lambda)(s,R(s))}E(R,W,F)R^{-1}(s). 
\end{eqnarray*} 
This inequality shows that it is sufficient to bound $e^{\mu}E.$ By
using the characteristic equation for $E$ in
(\ref{charE}) and inserting the expressions for $\mu_{r}$ and 
$\lambda_{t}$ and using the assumptions of the lemma it is an
immediate fact that $$\frac{d}{ds}(e^{\mu(s,R(s))}E)\leq 
0.$$   
\begin{flushright} 
$\Box$ 
\end{flushright} 
We now state and prove the main result of this section. 
\begin{theorem}Let $f_{0}\in I_{R_{-},F_{-}}$ and let
$(R(s,0,r,w,F),W(s,0,r,w,F),F)$ be any solution to the characteristic
system (\ref{char1})-(\ref{char2}) on $[0,T[$ such that 
$f_{0}(r,w,F)\not= 0,$ and assume that (\ref{hypmur}) and (\ref{hypmur-lr}) hold along this
  characteristic. Then $Q(t)$ is bounded on $[0,T[.$ 
\end{theorem} 
\textit{Proof. }We can without loss of generality assume that
$F_{-}\geq 1.$ The argument gets more transparent with this assumption
but it is clear from below that it is easy to modify the argument for
any $F_{-}>0.$ Let $t_{0}=T-1/6.$ Since the solution is regular on
$[0,T[$ we have that $C_{0}:=Q(t_{0})<\infty,$ in particular
$R(t_{0})\geq\frac{\sqrt{F_{-}}}{C_{0}},$ and $|W|\leq C_{0},$ for all
characteristics $(R(t),W(t),F).$ We now consider the time interval
$[t_{0},T[$ and we show that all matter stays uniformly
away from the centre on $[t_{0},T[$. The class of characteristics with
$R(t_{0})\geq 1/4$ will have $R(t)\geq 1/12$ for all $t\in[t_{0},T[$
by the fact that $|dR/ds|\leq 1.$ Hence we consider the complementary
class of characteristics with $\frac{\sqrt{F_{-}}}{C_{0}}\leq R(t_{0})\leq
  1/4.$ First we note that if $dR/ds>0$ then $dR/ds\neq 0,$
  i.e. $W\neq 0,$ as long as 
  $R\leq 1/3$ since we have for any $t$ with $W(t)=0$ and $R(t)\leq 1/3,$ 
\begin{eqnarray}
&\displaystyle\frac{dW}{ds}=-e^{\mu-\lambda}\mu_{r}(1+F/R^{2})+\frac{Fe^{\mu-\lambda}}{R^{3}E}&\nonumber\\
&\displaystyle\geq\left(\frac{-1}{R}+\frac{1}{R\log{(1/R)}}+\frac{F}{R^{3}\log{(1/R)}}\right)e^{\mu-\lambda}>0,&\nonumber 
\end{eqnarray}
by (\ref{hypmur}) and since $F\geq 1.$ 
Thus by continuity, an outgoing characteristic cannot turn around and
become ingoing 
as long as $R\leq 1/3.$ If it turns back when $R\geq 1/3$ it can
never get closer than $R=1/4$ for $t\leq T.$ 
We have thus found that the case of
interest is when $dR/ds<0$ and $\frac{\sqrt{F_{-}}}{C_{0}}\leq
  R(t_{0})\leq 1/4.$ We now show that for
any such characteristic, $W$ will become zero before $R$ becomes zero. 
By (\ref{char2}) we get 
\begin{eqnarray}
&\displaystyle\frac{d}{ds}(e^{2\lambda}W^2)=-2\lambda_{r}\frac{W}{E}e^{\mu+\lambda}W^{2}-2\frac{W}{E}e^{\mu+\lambda}\mu_{r}(1+W^{2}+F/R^{2})+\frac{2WFe^{\mu+\lambda}}{R^{3}E}&\nonumber\\
&\displaystyle
=-2(\mu_{r}-\lambda_{r})\frac{dR}{ds}e^{2\lambda}W^{2}-\frac{2F}{R^{2}}(\mu_{r}-\frac{1}{R})\frac{dR}{ds}e^{2\lambda}-2\frac{dR}{ds}e^{2\lambda}\mu_{r}.&\nonumber 
\end{eqnarray}
Hence, by (\ref{hypmur}), and the fact that
$\log{R}<-1$ and $F\geq 1$ we have 
\begin{eqnarray}
&\displaystyle\frac{d}{ds}\left(e^{\int_{0}^{s}2(\mu_{r}-\lambda_{r})\frac{dR}{ds}d\tilde{s}}e^{2\lambda}W^2\right)&\nonumber\\
&\displaystyle =\left(-\frac{2F}{R^{2}}(\mu_{r}-\frac{1}{R})\frac{dR}{ds}e^{2\lambda}-2\frac{dR}{ds}e^{2\lambda}\mu_{r}\right)e^{\int_{0}^{s}2(\mu_{r}-\lambda_{r})\frac{dR}{ds}d\tilde{s}}&\nonumber\\
&\displaystyle\leq\left(\frac{-2F}{R^{3}\log{R}}-2\mu_{r}\right)e^{2\lambda}\frac{dR}{ds}e^{\int_{0}^{s}2(\mu_{r}-\lambda_{r})\frac{dR}{ds}d\tilde{s}}&\nonumber\\
&\displaystyle\leq\frac{-F}{R^{3}\log{R}}\frac{dR}{ds}e^{\int_{0}^{s}2(\mu_{r}-\lambda_{r})\frac{dR}{ds}d\tilde{s}}&\nonumber\\
&\displaystyle\leq\frac{-F}{R^{3}\log{R}}\frac{dR}{ds}R^{2}\leq\frac{-F}{R\log{R}}\frac{dR}{ds}.&
\end{eqnarray}
Integrating this inequality gives 
\begin{eqnarray}
&\displaystyle e^{\int_{t_{0}}^{t}2(\mu_{r}-\lambda_{r})\frac{dR}{ds}ds}e^{2\lambda(t,R(t))}W^{2}(t)-e^{2\lambda(t_{0},R(t_{0}))}W^{2}(t_{0})\leq&\nonumber\\
&\displaystyle
\leq -F\log{|\log{R(t)}|}+F\log{|\log{R(t_{0})}|},& 
\end{eqnarray}
so that 
\begin{eqnarray}
&\displaystyle e^{2\lambda(t,R(t))}W^{2}(t)\leq&\nonumber\\
&\displaystyle\leq
e^{-\int_{t_{0}}^{t}2(\mu_{r}-\lambda_{r})\frac{dR}{ds}ds}\left(e^{2\lambda(t_{0},R(t_{0}))}W^{2}(t_{0})-F\log{\frac{|\log{R(t)}|}{|\log{R(t_{0})}|}}\right).& 
\end{eqnarray}
This relation implies that $W$ will vanish before $R$ becomes $0.$
From the arguments of the first part of the proof we have seen that
$dW/ds>0$ when $W=0,$ so that the characteristic necessarily becomes
outgoing and cannot turn back in again as long as $R\leq 1/3.$ 
\begin{flushright}
$\Box$
\end{flushright}

\section{The case of purely ingoing matter} 
A special case of interest is when all matter is falling inwards. This
is for example the situation when dust is used as matter
model; if the matter is initially falling inwards, there will be no
outgoing matter at any time. Moreover, in the numerical
simulation~\cite{RRS2} most matter 
is seen to be falling inwards in the supercritical case when black
holes (presumably) form. We believe that a good understanding of this
case will also have impact on the general case since control of
ingoing matter give (in a certain sense) control of the outgoing part. 
In this section we will take initial data 
such that all particles are initially falling inwards and we will
assume that all matter continue to fall inwards at least on some time
interval $[0,T[.$ The other assumption is connected to the relation
between the enclosed mass $m(t,r)$ and the areal radius $r.$
It is well-known that there are closed null geodesics when $r=3M$ in 
the Schwarzschild spacetime with mass $M.$ 
Therefore it is not surprising that the relation $r=3m$ turns out to
have significance also in the Einstein-Vlasov system. We will show
that when all matter is falling inwards, control of ingoing
characteristics can be obtained either if $r\leq 3m(t,r)$ everywhere, or if
$r\geq 3m$ along the characteristics. Before stating our 
hypotheses we define the outgoing and ingoing part of $j,$ 
\begin{eqnarray}
j^{+}(t,r)&=&\frac{\pi}{r^{2}}
\int_{0}^{\infty}dw\int_{0}^{\infty}dF\,wf(t,r,w,F),\\ 
j^{-}(t,r)&=&\frac{\pi}{r^{2}}
\int_{-\infty}^{0}dw\int_{0}^{\infty}dF\,wf(t,r,w,F). 
\end{eqnarray} 
On the time interval $[0,T[$ we assume that 
\begin{eqnarray}
&j^{+}=0,&\label{hypinw}\\
&r\geq 3m(t,r).&\label{hypr3m} 
\end{eqnarray}
We remark that the first hypothesis can be replaced by assuming 
that $j^{+}$ is bounded but we find it more transparent to think of
the results in this section as the case of purely ingoing matter. 
The first question we ask is what can happen if the second hypothesis
is not satisfied and the following lemma gives a partial answer. 
\begin{lemma}
Let $(R(s),W(s),F)$ be any solution to the characteristic system on a
time interval $I=[t_{1},t_{2}]$ such that $R(t_{1})\geq\delta>0.$
Assume that (\ref{hypinw}) holds and that 
\begin{displaymath}
R(s)\leq 3m(s,R(s)), 
\end{displaymath} 
on $I.$ Then there exists an $\epsilon>0,$ which only depends on
$\delta$ such that 
\[
R(s)\geq\epsilon,\; s\in[t_{1},t_{2}]. 
\]
%\begin{displaymath}
%\frac{1}{R(s)}\leq\beta(s), 
%\end{displaymath} 
%where $\beta\in C([t_{1},t_{2}])$ depends only on $\delta.$ 
\end{lemma} 
\textit{Proof. }The proof is similar to the proof of Lemma 5. Indeed, 
recall that 
\begin{equation} 
\mu(t,r)=-\int_{r}^{\infty}e^{2\lambda}(\frac{m(t,\eta)}{\eta^{2}}+4\pi
\eta p(t,\eta))d\eta.\label{mu} 
\end{equation} 
Let us write 
\begin{equation} 
\mu=\hat{\mu}+\check{\mu},\label{musplit} 
\end{equation} 
where 
\begin{displaymath} 
\hat{\mu}(t,r)=-\int_{r}^{\infty}e^{2\lambda}\frac{m(t,\eta)}{\eta^{2}}d\eta. 
\end{displaymath} 
Since $e^{\hat{\mu}}\leq e^{\mu}$ we proceed as in Lemma 5 and
consider $\frac{d}{ds}(e^{\hat{\mu}}/R).$ From the equations (\ref{ee3}) and
(\ref{e2-lambda} ) it follows that 
\begin{equation} 
\hat{\mu}_{t}(t,r)=\int_{r}^{\infty} 4\pi
je^{\lambda+\mu}(1+2e^{2\lambda}\frac{m(t,\eta)}{\eta})d\eta,\label{mut} 
\end{equation} 
so $\mu_{t}\leq 0$ by (\ref{hypinw}). It is then easy to show that 
$e^{\hat{\mu}}/R$ is nonincreasing by a similar calculation as in
Lemma 5. 
\begin{flushright}
$\Box$
\end{flushright} 
The main result in this section concerns an ingoing characteristic
for which $d(e^{\hat{\mu}}E)/ds\geq 0.$ We have seen above that
control of this quantity is sufficient for controlling $R(t)^{-1}.$ 
The condition that the time derivative is nonnegative should then 
be the relevant case. One can 
think of dividing the time interval into subintervals where $e^{\hat{\mu}}E$
does increase and where it decreases and then add up the contributions
where $d(e^{\hat{\mu}}E)/ds\geq 0.$ However, if the sign of
$d(e^{\hat{\mu}}E)/ds$ changes an infinite number of times on $[0,T[,$
then it is not clear that the infinite sum of the contributions
converge due to the nature of the estimate we obtain below. This is
the reason for including this assumption. 
\begin{theorem} 
Let $f_{0}\in I_{R_{-},F_{-}}$ and such that $f_{0}(\cdot,w,\cdot)=0$
if $w>0.$ Let 
$(R(s,0,r,w,F),W(s,0,r,w,F),F)$ be any solution to the characteristic
system (\ref{char1})-(\ref{char2}) on $[0,T[$ such that
$f_{0}(r,w,F)\not= 0.$ Assume that $d(e^{\hat{\mu}}E)/ds\geq 0$ on
$[0,T[,$ and assume that (\ref{hypinw}) and (\ref{hypr3m}) hold. Then 
there exists an
$\epsilon>0$ such that 
\begin{equation} 
R(s,0,r,w,F)>\epsilon,\; \mbox{for all } s\in [0,T[. 
\end{equation} 
\end{theorem} 
\textit{Proof. }We will only sketch the proof since it is to some
extent similar to the proof of Theorem 1. Again we consider 
$e^{\hat{\mu}}E$ and we show that this quantity can be controlled. 
We have 
\begin{eqnarray*}
&\displaystyle\frac{d}{ds}\left( E(R(s),W(s),F)e^{\hat{\mu}(s,R(s))}\right)=&\\ &\displaystyle =4\pi\left[ Re^{\mu+\lambda}\left(
  \frac{W^{2}}{E^{2}}j+\frac{|W|}{E}p\right)+\int_{{R(s)}}^{\infty}
e^{\mu+\lambda}j\left(1+\frac{2e^{2\lambda}m}{\eta}\right) 
d\eta\right]Ee^{\hat{\mu}}.& 
\end{eqnarray*}
Here we have inserted the formula (\ref{mut}) for $\mu_{t}.$ Note also 
that the terms involving $m/R^{2}$ have cancelled. 
From the assumption that $e^{\hat{\mu}}E$ is nondecreasing we have
(note that the integral term is nonpositive by (\ref{hypinw})) 
$$\frac{W^{2}}{E^{2}}j+\frac{|W|}{E}p\geq 0,$$ 
and thus 
$$\frac{W^{2}}{E^{2}}j+\frac{|W|}{E}p\leq \frac{|W|}{E}j+p.$$ 
We get 
\begin{eqnarray} 
&\displaystyle\frac{d}{ds}\left(E(R(s),W(s),F)e^{\hat{\mu}(s,R(s))}\right)&\nonumber\\ 
&\displaystyle\leq 4\pi\left[Re^{\mu+\lambda}\left(
  \frac{|W|}{E}j+p\right)+\int_{{R(s)}}^{\infty}
e^{\mu+\lambda}j(s,\eta)\left(1+\frac{2e^{2\lambda}m(s,\eta)}{\eta}\right)
d\eta
\right]Ee^{\hat{\mu}}.&\nonumber\\
\label{charehatmuE} 
\end{eqnarray} 
This is a similar to equation (\ref{charE}) and we are again going to
apply Green's formula in the $(t,r)-$plane. This time we will choose
the \textit{outer} domain $\Omega,$ which is enclosed by the curve
$(t,R(t)),\; t\in [0,T[,$ which we denote by $\gamma,$ together with
the curves 
\begin{eqnarray*} 
C_{2}&=&\{(T,r):R(T)\leq r\leq R^{\infty}\},\\ 
C_{3}&=&\{(t,R^{\infty}):T\geq t\geq 0\},\\ 
C_{4}&=&\{(0,r):R^{\infty}\geq r\geq R(0)\}. 
\end{eqnarray*} 
Here $R^{\infty}\geq R_{+}+T,$ so that $f=0$ when $r\geq R^{\infty}.$ 
The closed curve $\gamma +C_{2}+C_{3}+C_{4},$ we denote by $C$ and it
is oriented clockwise. 
We integrate (\ref{charehatmuE}) in time and obtain a curve integral
over $\gamma,$ analogous to (\ref{curve1}), and we note that the part 
containing $|W|j/E+p$ can be written 
$$\displaystyle\int_{\gamma} e^{-\mu+\lambda}\lambda_{t}dr+
e^{\mu-\lambda}\check{\mu}_{r}ds,$$ and we then have by Green's
formula 
\begin{eqnarray} 
&\displaystyle\oint_{C} e^{-\mu+\lambda}\lambda_{t}dr+
e^{\mu-\lambda}\check{\mu}_{r}ds&\nonumber\\ 
&\displaystyle =\int\int_{\Omega}\partial_{t}\left(e^{-\mu+\lambda}
\lambda_{t}\right)-\partial_{r}
\left( e^{\mu-\lambda}\check{\mu}_{r}\right)dsdr&\nonumber\\ 
&\displaystyle =\int\int_{\Omega}\partial_{t}\left(e^{-\mu+\lambda}
\lambda_{t}\right)-\partial_{r}
\left( e^{\mu-\lambda}\mu_{r}\right)dsdr 
+\int\int_{\Omega}\partial_{r}
\left(e^{\mu+\lambda}\frac{m}{r^{2}}\right)dsdr.&\nonumber\\ 
\end{eqnarray}
Using the calculations that led to equation (\ref{id}) we obtain
the identity 
\begin{eqnarray}
&\displaystyle\oint_{C} e^{-\mu+\lambda}\lambda_{t}dr+
e^{\mu-\lambda}\check{\mu}_{r}ds&\nonumber\\
&\displaystyle=\int\int_{\Omega}4\pi e^{\mu+\lambda}\left[ 
(\rho+p)e^{2\lambda}\frac{m}{r}+2p-\rho\right] drds&\nonumber\\ 
&\displaystyle+\int\int_{\Omega}\int_{-\infty}^{\infty}\int_{0}^{\infty}
\frac{4\pi^{2} e^{\mu+\lambda}}{r^{2}E}f(t,r,w,F)dFdw\,drds.&\label{id2} 
\end{eqnarray} 
Recall from (\ref{charehatmuE}) that we wish to compute the curve integral
\begin{eqnarray} 
&\displaystyle\int_{0}^{T}4\pi Re^{\mu+\lambda}\left(
  \frac{|W|}{E}j+p\right)+\int_{0}^{T}\int_{R(s)}^{\infty} 4\pi
je^{\lambda+\mu}(1+2e^{2\lambda}\frac{m}{\eta})d\eta ds&\nonumber\\ 
&\displaystyle =\oint_{C} e^{-\mu+\lambda}\lambda_{t}dr+
e^{\mu-\lambda}\check{\mu}_{r}ds+\int_{0}^{T}\int_{R(s)}^{\infty} 4\pi
je^{\lambda+\mu}(1+2e^{2\lambda}\frac{m}{\eta})d\eta ds&\nonumber\\ 
&\displaystyle -\int_{C_{2}+C_{3}+C_{4}} e^{-\mu+\lambda}\lambda_{t}dr+
e^{\mu-\lambda}\check{\mu}_{r}ds.&\label{curveC234} 
\end{eqnarray} 
Since 
$j=0$ when $r\geq R^{\infty}$ we get 
\begin{eqnarray}
&\displaystyle\oint_{C} e^{-\mu+\lambda}\lambda_{t}dr+
e^{\mu-\lambda}\check{\mu}_{r}ds+\int_{0}^{T}\int_{R(s)}^{\infty} 4\pi
je^{\lambda+\mu}(1+2e^{2\lambda}\frac{m}{\eta})d\eta ds&\nonumber\\
&\displaystyle=\int\int_{\Omega}4\pi e^{\mu+\lambda}\left[ 
(\rho+p)e^{2\lambda}\frac{m}{r}+2p-\rho+
j(1+2e^{2\lambda}\frac{m}{r})\right] drds&\nonumber\\ 
&\displaystyle+\int\int_{\Omega}\int_{-\infty}^{\infty}\int_{0}^{\infty}
\frac{4\pi^{2} e^{\mu+\lambda}}{r^{2}E}f(t,r,w,F)dFdw\,drds.&\label{green3} 
\end{eqnarray}
The definitions of the matter terms and assumption (\ref{hypinw}) imply that 
\begin{equation} 
p\leq -j\leq\rho,\label{p<|j|<rho} 
\end{equation} 
and the first term in (\ref{green3}) can now be estimated by 
\begin{eqnarray*} 
&\displaystyle(\rho+p)e^{2\lambda}\frac{m}{r}+2p-\rho+
j(1+2e^{2\lambda}\frac{m}{r})&\nonumber\\
&\displaystyle =
-(\rho-p)(1-e^{2\lambda}\frac{m}{r})+(p+j)(1+2e^{2\lambda}\frac{m}{r})\leq
0,&\nonumber 
\end{eqnarray*} 
where we used 
$$e^{2\lambda}\frac{m}{r}=\frac{m/r}{1-2m/r}\leq 1,$$ 
by assumption (\ref{hypr3m}). 
A bound for the second term in (\ref{green3}) follows as in the proof
of Theorem 1 where $T_{1}^{prin}$ was introduced. 
The two first terms in (\ref{curveC234}) have now been estimated and
we are left with the boundary term 
\begin{displaymath} 
 -\int_{C_{2}+C_{3}+C_{4}} e^{-\mu+\lambda}\lambda_{t}\,dr+
e^{\mu-\lambda}\check{\mu}_{r}\,ds. 
\end{displaymath} 
The integral over $C_{2}$ is nonpositive by assumption (\ref{hypinw}) since 
\begin{displaymath} 
-\int_{C_{2}}e^{-\mu+\lambda}\lambda_{t}\,dr+
e^{\mu-\lambda}\check{\mu}_{r}\,ds=-\int_{R(T)}^{R^{\infty}}
e^{-\mu+\lambda}\lambda_{t}\,dr=\int_{R(T)}^{R^{\infty}}
4\pi e^{2\lambda}rj\leq 0. 
\end{displaymath} 
The integral over $C_{3}$ is zero by letting $R^{\infty}$ tend to
infinity since $m\leq M.$ Finally, the integral
over $C_{4}$ depends only on the initial data and is thus bounded (we 
remark that it is at this point it is troublesome to add up an
infinite sum of contributions as was previuosly discussed) 
which completes the proof of the theorem. 
\begin{flushright} 
$\Box$ 
\end{flushright} 
\textit{Remark. }An alternative way of presenting a similar result
would have been to assume that the sign of $d(e^{\hat{\mu}}E)/ds,$ and
of $R(t)-3m(t,R(t))$ only change a finite number of times on $[0,T[.$ These
assumptions together with (\ref{hypinw}) would be sufficient for
global existence. Such a proof would rest on the results in this
section together with a more involved analysis of the boundary terms. 
\begin{center}
\textbf{Acknowledgement} 
\end{center} 
The author wants to thank Markus Kunze for stimulating discussions.

\end{document}